\documentclass[aps,prd,onecolumn,a4paper,superscriptaddress,nofootinbib]{revtex4-1}
\usepackage{tabularx, amsmath, amssymb, graphicx, multirow, dcolumn, bm, latexsym, soul, nicefrac, subfigure, graphicx, appendix,acronym}
\usepackage{mathrsfs, amsthm, amssymb, enumerate, bm}
\usepackage[normalem]{ulem}
\usepackage[colorlinks, linkcolor=blue, citecolor=blue, urlcolor=blue ]{hyperref}
\usepackage{ulem} % used solely for the purpose of enable strike-through lines. Should remove/comment before submission
\begin{document}
%\begin{CJK*}{GB}{} % CJK
%%%%%%%%%%%%%%%%%%%%%%%%%%%%%%%%%%%%%%%%%%%%%%%%%%%%%%%%%%%%%%%%
\newcommand{\ls}[1]{\textcolor{blue}{LS: #1}}
\newcommand{\yh}[1]{\textcolor{red}{YH: #1}}
\newcommand{\lz}[1]{\textcolor{cyan}{\sf{#1}}}
\newcommand{\mnras}{Monthly Notices of the Royal Astronomical Society}
\newcommand{\apjl}{the Astrophysical Journal Letters}
\newcommand{\apjs}{the Astrophysical Journal Supplement}
\newcommand{\araa}{Annual Review of Astronomy and Astrophysics}
\newcommand{\aapr}{Astronomy and Astrophysics Reviews}
\newcommand{\pasa}{Publications of the Astronomical Society of Australia}
\newcommand{\aap}{Astronomy and Astrophysics}
\newcommand{\msun}{M_\odot}
\newcommand{\mbh}{M_{\rm BH}}
\newcommand{\mpbh}{M_{\rm PBH}}
\newcommand{\Sa}{\,{\rm m}\,{\rm s}^{-2}\,{\rm Hz}^{-1/2}}
\newcommand{\Sx}{\,{\rm m}\,{\rm Hz}^{-1/2}}
\newcommand{\D}{\mathrm{d}}
\def\ccr#1{{\color{red}{\bf #1}}}
%%%%%%%%%%%%%%%%%%%%%%%%%%%%%%%%%%%%%%%%%%%%%%%%%%%%%%%%%%%%%%%%
\title{Capability for detection of GW190521-like binary black holes with TianQin}

  \author{Shuai Liu}
\email{liushuai5@mail.sysu.edu.cn}
\affiliation{MOE Key Laboratory of TianQin Mission, TianQin Research Center for Gravitational Physics $\&$ School of Physics and Astronomy, Frontiers Science Center for TianQin, Gravitational Wave Research Center of CNSA, Sun Yat-sen University (Zhuhai Campus), Zhuhai 519082,  People's Republic of China}
  
  \author{Liang-Gui Zhu}
\affiliation{MOE Key Laboratory of TianQin Mission, TianQin Research Center for Gravitational Physics $\&$ School of Physics and Astronomy, Frontiers Science Center for TianQin, Gravitational Wave Research Center of CNSA, Sun Yat-sen University (Zhuhai Campus), Zhuhai 519082,  People's Republic of China}

  \author{Yi-Ming Hu}
\email{huyiming@sysu.edu.cn}
\affiliation{MOE Key Laboratory of TianQin Mission, TianQin Research Center for Gravitational Physics $\&$ School of Physics and Astronomy, Frontiers Science Center for TianQin, Gravitational Wave Research Center of CNSA, Sun Yat-sen University (Zhuhai Campus), Zhuhai 519082,  People's Republic of China}
  
  \author{Jian-dong Zhang}
\affiliation{MOE Key Laboratory of TianQin Mission, TianQin Research Center for Gravitational Physics $\&$ School of Physics and Astronomy, Frontiers Science Center for TianQin, Gravitational Wave Research Center of CNSA, Sun Yat-sen University (Zhuhai Campus), Zhuhai 519082,  People's Republic of China}
  
  \author{Mu-Jie Ji}
\affiliation{MOE Key Laboratory of TianQin Mission, TianQin Research Center for Gravitational Physics $\&$ School of Physics and Astronomy, Frontiers Science Center for TianQin, Gravitational Wave Research Center of CNSA, Sun Yat-sen University (Zhuhai Campus), Zhuhai 519082,  People's Republic of China}
\affiliation{Shantou Jinshan Middle School, Shantou 515073, People's Republic of China}

\date{\today}

%%%%%%%%%%%%%%%%%%%%%%%%%%%%%%%%%%%%%%%%%%%%%%%%%%%%%%%%%%%%%%%%
\begin{abstract}
The detection of GW190521 gains huge attention because it is the most massive binary that LIGO and Virgo ever
    confidently detected until the release of GWTC-3 (GW190426\_190642 is more massive), and it is the first black hole merger whose remnant is believed to be an intermediate mass black hole.
    Furthermore, the primary black hole mass falls in the black hole mass gap, where the pair-instability supernova
    prevents the formation of astrophysical black holes in this range. In this paper, we systematically explore
    the prospect of TianQin on detecting GW190521-like sources. For sources with small orbital eccentricities, (i)
    TianQin could resolve up to a dozen of sources with signal-to-noise ratio (SNR) larger than 8. Even if the
    signal-to-noise ratio 
    threshold increases to 12, TianQin could still detect GW190521-like binaries. (ii) The parameters of sources merging
    within several years would be precisely recovered. The precision of coalescence time and sky localization closes to
    $1\ {\rm s}$ and $1\ {\rm deg^{2}}$ respectively.
This indicates that TianQin could provide early warnings for ground-based gravitational waves detectors and electromagnetic telescopes for these sources. Furthermore, TianQin could distinguish the formation channels of these sources by
    measuring the orbital eccentricities with a relative precision of $10^{-4}$.
  (iii) TianQin could constrain the Hubble constant with a $10\%$ precision with GW190521-like sources.
  Finally, for very eccentric GW190521-like sources, although their gravitational wave signal might be too weak for TianQin to detect,
    even the null detection of TianQin could still present a significant contribution to the understanding of the underlying science. 
\end{abstract}
%%%%%%%%%%%%%%%%%%%%%%%%%%%%%%%%%%%%%%%%%%%%%%%%%%%%%%%%%%%%%%%%
\maketitle
%\end{CJK*} % CJK
%%%%%%%%%%%%%%%%%%%%%%%%%%%%%%%%%%%%%%%%%%%%%%%%%%%%%%%%%%%%%%%%
%%%%%%%%%%%%%%%%%%%%%%%%%%%%%%%%%%%%%%%%%%%%%%%%%%%%%%%%%%%%%%%%
\section{Introduction}
During the first three observations of Laser Interferometer Gravitational-Wave Observatory (LIGO)
and Virgo \cite{TheLIGOScientific:2014jea, VIRGO:2014yos}, more than 90 gravitational wave (GW) events have been reported so far
\cite{LIGOScientific:2018mvr, LIGOScientific:2020ibl, LIGOScientific:2021qlt, LIGOScientific:2021usb,
LIGOScientific:2021djp}, of which the event GW190521 \cite{LIGOScientific:2020iuh, LIGOScientific:2020ibl}
drew great attention. 
This is because it is more massive than any other system detected before it (note that GW190426\_190642 becomes the most massive event after GWTC-3 is released), and it is the first stellar-mass binary black hole (SBBH) event whose
primary mass ($\bm{95.3_{-18.9}^{+28.7}}M_{\odot}$) falls in the mass gap, and whose remnant mass
$\bm{163.9_{-23.5}^{+39.2}}M_{\odot}$ is considered to be an intermediate mass black hole (IMBH).

In the mass spectrum of stellar-mass black holes (SBHs) from the stellar evolution, it is predicted that a mass gap of
about $65-120M_{\odot}$ would occur, due to the process known as the pair-instability supernova
\cite{Heger:2001cd, Belczynski:2016jno, Woosley:2016hmi, Spera:2017fyx, Marchant:2018kun}. If SBHs are generated
by stellar collapses, then the custom wisdom predicts no existence of black holes within the mass gap. The detection of
GW190521 triggered huge interest in the studying of its formation mechanism, which could be roughly divided into two
categories: (i) stellar evolution from isolated binaries with zero or very low metallicity \cite{Vink:2020nak, Tanikawa:2020abs,
Costa:2020xbc, Farrell:2020zju} (the stars with low metallicity could keep most of their hydrogen envelope until the
presupernova phase, avoid the pair-instability supernova explosions and produce SBHs within the mass gap by fallback),
and (ii) pairing through dynamical processes.
The SBH falling in the mass gap could be formed by mergers of SBHs/stars, i.e., hierarchical mergers, \cite{Fragione:2020han, Martinez:2020lzt, Samsing:2020qqd, Anagnostou:2020umw, Gerosa:2021mno, Wang:2021mdt,Mapelli:2021syv, Tagawa:2020qll, Kimball:2020qyd, Liu:2020gif, DiCarlo:2019pmf, DiCarlo:2019fcq,Kremer:2020wtp, Renzo:2020smh} or accretion onto SBHs \cite{Safarzadeh:2020vbv, Natarajan:2020avl, Liu:2020lmi, Cruz-Osorio:2021qbr, Rice:2020gyx}, and then form a SBBH by gravitational interaction with another SBH. 
It is expected that the orbital eccentricities would be used to decipher the formation channels. 
The SBBHs formed through the binary evolution are expected to have circular orbits, while the ones formed by the
dynamical process would have measurable eccentricties \cite{Samsing:2013kua, Antonini:2015zsa, Nishizawa:2016jji, Breivik:2016ddj, Chen:2017gfm, LIGOScientific:2018jsj, Samsing:2018isx, Kremer:2018cir}. 
Several exotic formation channels are also proposed \cite{Carr:2019kxo, DeLuca:2020sae,Clesse:2020ghq, Liu:2021jdz, Sakstein:2020axg, Palmese:2020xmk, Antoniou:2020syc, CalderonBustillo:2020srq}, e.g., primordial BHs, modified gravity. Finally, the probability that GW190521 is a SBBH straddling the mass gap is proposed \cite{Fishbach:2020qag}.

Intermediate mass black holes are black holes with masses between $10^{2}-10^{5}M_{\odot}$, and they are speculated to exist in the center of dwarf galaxies or globular clusters (e.g. \cite{Miller:2003sc, vanderMarel:2003ka}). 
Since their masses locate between those of SBHs and of supermassive black holes (SMBHs), their discovery is expected to shed light on the formation of SMBHs \cite{Volonteri:2010wz, Greene:2019vlv}.
There are many different channels that could produce IMBHs, for example, the collapse of population III stars or gas clouds with low angular momentum \cite{Fryer:2000my, Heger:2002by, Spera:2017fyx, Loeb:1994wv, Bromm:2002hb, Lodato:2006hw}, mergers of SBHs \cite{Miller:2001ez, OLeary:2005vqo, 2015MNRAS.454.3150G}, and runaway collisions of stellar objects \cite{PortegiesZwart:2002iks, AtakanGurkan:2003hm, PortegiesZwart:2004ggg}. 
However, previous detections of IMBHs were made through indirect observations.
For example, there are several candidates reported as ultraluminous x-ray sources (e.g.,\cite{Mezcua:2015pra,
Mezcua:2017npy, Wang:2015sma, Kaaret:2017tcn, Lin:2020exl}), and some are reported as associated with low luminous
active galactic nuclei (e.g., \cite{Baldassare:2016jit, Baldassare:2016cox, Mezcua:2018ifv}). 
The successful observation for GW190521 has made a breakthrough in the search for IMBHs. 
This is because the GW observation provides a direct measurement of the remnant mass, which confirms the existence of the merger channel for IMBH formation.

Although GW190521 was observed by LIGO/Virgo at $\sim 100\ {\rm Hz}$, the early inspiral GWs from such binaries
could also be detectable in the lower frequencies. It has been shown that space-borne detectors, such as Laser
Interferometer Space Antenna (LISA) \cite{LISA:2017pwj}, could not only detect such systems but also measure the environment effects through their effects on waveforms \cite{Toubiana:2020drf}.
Furthermore, by observing GW190521-like events with space-borne GW detectors, it is possible to distinguish the
formation channels by measuring the orbital eccentricities \cite{Mandel:2017pzd, Holgado:2020imj}, test general relativity (GR) by
constraining the GR deviation parameters \cite{Mastrogiovanni:2020mvm}, and measure the Hubble constant by
treating the system as a standard siren \cite{Chen:2020gek, Mukherjee:2020kki, Mastrogiovanni:2020mvm, Gayathri:2021isv}.

TianQin is a space-borne GW observatory that is planned to be launched in the 2030s \cite{Luo:2015ght}.
The shorter arm length with TianQin supports a better sensitivity for higher frequencies, making it sensitive to the early inspirals of SBBHs.
It has been shown that TianQin could detect up to dozens of SBBHs and recover their parameters very accurately
\cite{Liu:2020eko}, and these future detections could be used to constrain the Hubble constant to a precision of
$\sim1\%$ in the most ideal case \cite{Zhu:2021bpp}. 
In this work, we will focus on the capability of TianQin for inspiral GWs from GW190521-like binaries, and how the
future detections could be used to constrain cosmology.

The rest of this paper is organized as follows. 
In Sec. \ref{Sec:DetectionCapability}, we estimate the detection number and parameter estimation precision for GW190521-like binaries with small eccentricities.
Depending on these results, we explore the potential of TianQin to provide early warning for ground-based GW
detectors/EM telescopes and distinguish the formation channels by measuring the orbital eccentricities.
We explore the application of such detections on GW cosmology in Sec. \ref{Sec:ConstrainHubbleConstant}.
In Sec. \ref{Sec:Discussion}, we shift our attention to the orbital eccentricities and assess the potential of
TianQin to find very eccentric GW190521-like binaries through archival search, assuming a joint observation of TianQin and the future generation ground-based GW detectors.
We also discuss the possibility that GW190521-like binaries are intrinsically light systems that appear much heavier due
to environmental effects. 
We draw the conclusion in Sec. \ref{Sec:Conclusion}.
Throughout the paper, we use the geometrical units ($G=c=1$) and masses in the source rest frame unless otherwise specified.
Furthermore, we adopt the standard $\Lambda {\rm CDM}$ cosmological model \cite{Planck:2015fie}.

%%%%%%%%%%%%%%%%%%%%%%%%%%%%%%%%%%%%%%%%%%%%%%%%%%%%%%%%%%%%%%%%
\section{The detection capability of TianQin}\label{Sec:DetectionCapability}

\subsection{Detection number}
The frequency band where TianQin is most sensitive is $\sim10^{-2}{\rm Hz}$.
At this frequency, the orbital eccentricities of SBBHs formed by the dynamical process are predicted to be accompanied
with $e \sim 0.001-0.1$, and the SBBHs from isolated binary evolution are expected to be associated with lower eccentricities \cite{Nishizawa:2016jji, Breivik:2016ddj, Chen:2017gfm}.
Therefore, in this section, we carry out the following calculation assuming that GW190521-like binaries have these small eccentricities.
This assumption simplifies the calculation as we can focus on the dominant $n=2$ harmonic \cite{Peters:1963ux, Peters:1964zz, Chen:2017gfm}.
It has been suggested that for space-borne detectors, the second order post-Newtonian (2PN) waveform is sufficiently
accurate, in a sense that the waveform systematic error is less than the statistic error \cite{Mangiagli:2018kpu}. 
Consequently, we adopt a 3PN waveform with eccentricity \cite{Krolak:1995md, Buonanno:2009zt, Feng:2019wgq} throughout the work.
The detectors we consider are as follows:
\begin{enumerate}[i.]
  \item TianQin: a regular triangle shaped space-borne detector follows a geocentric orbit. It observes in a ``3months
      on + 3months off" scheme, which would cause gaps in the record data.
    In addition to the fiducial one constellation (TQ) configuration, we also consider the twin constellation (TQ I+II)
        configuration to remove the gaps \cite{Liu:2020eko}.
        We adopt the sensitivity curve is from \cite{Wang:2019ryf}, and assume a fiducial operation time of five
        years.
        We do not consider the foreground noise from double white dwarfs (DWDs) throughout this work, as studies
        suggested that the influence of the foreground noise over 5 years is trivial for TianQin \cite{Huang:2020rjf, Liang:2021bde}.
  \item LISA: a regular triangle shaped space-borne detector follows a heliocentric orbit.
      We adopt the sensitivity curve with foreground noise from DWDs from \cite{Robson:2018ifk}, and assume a fiducial
        operation time of four years.
  \item LIGO A+: a right angle shaped ground-based detector.
    We adopt the power spectral density (PSD) from LIGO
        documents.\footnote{\href{https://dcc.ligo.org/LIGO-T1800042/public}{https://dcc.ligo.org/LIGO-T1800042/public}}
  \item Cosmic Explore (CE): a right angle shaped ground-based detector.
    We adopt the PSD from \cite{LIGOScientific:2016wof} and assume a network of two CE detectors as proposed by the CE team \cite{Reitze:2019iox}.
  \item Einstein Telescope (ET): a regular triangle shaped ground-based detector.
    We adopt the PSD from the ET-D configuration \cite{Hild:2010id}.
    Notice that unlike TianQin or LISA, ET has three independent interferometers.
\end{enumerate}

\begin{figure}[h]
\centering
\includegraphics[width=0.5\textwidth]{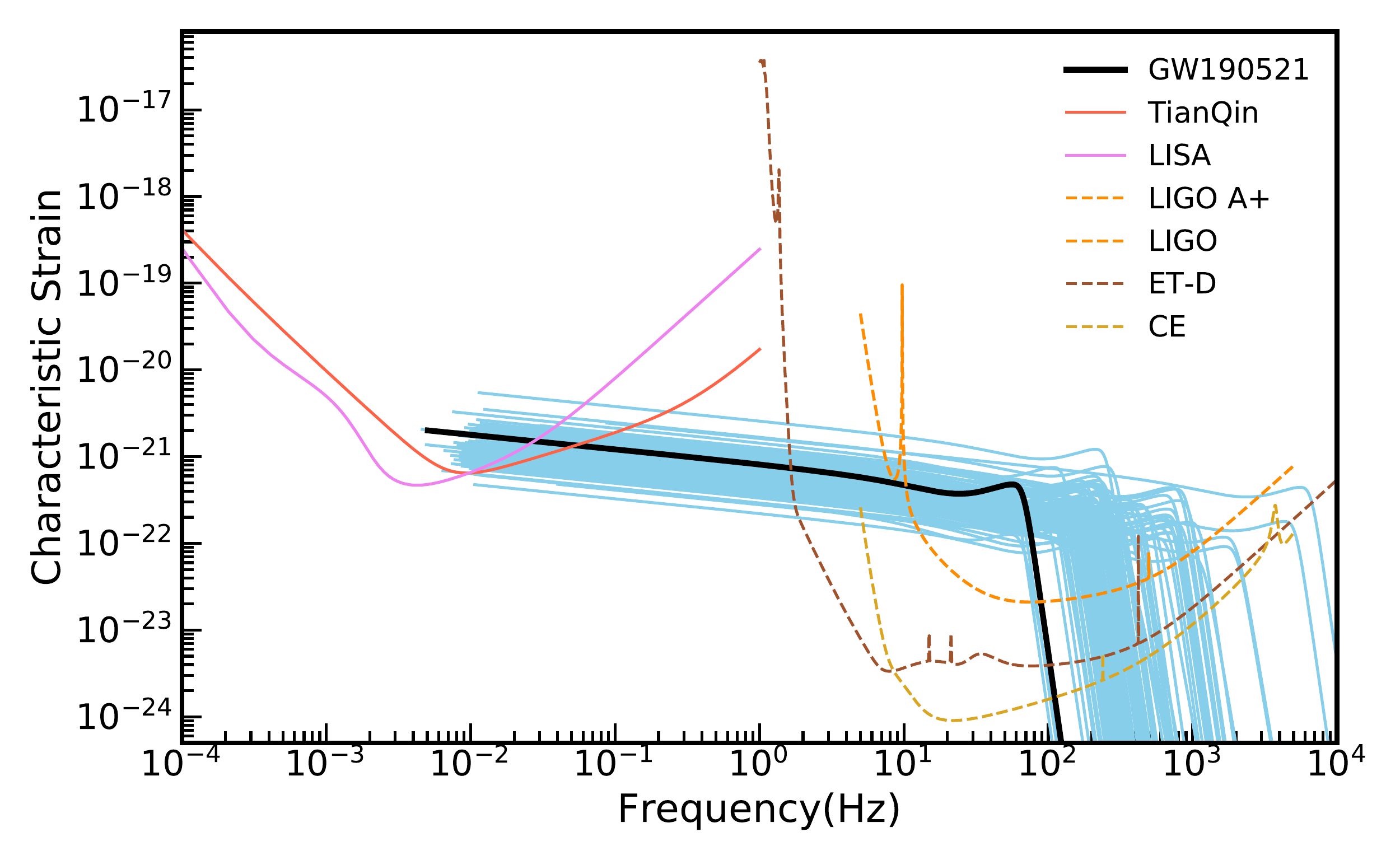}
    \caption{The GW characteristic amplitudes $2f|\tilde{h}(f)|$ of events which are assumed to be quasicircular
    from GWTC-1, GWTC-2, and GWTC-3 versus noise strains $\sqrt{fS_{n}(f)}$ of detectors. The notations $\tilde{h}(f)$ and
    $f$ are the GW signal in frequency domain and the corresponding GW frequency respectively, and $S_{n}(f)$ is the PSD of detector.
    The black solid curve indicates GW190521, and the blue solid curves represent the other events. The starting frequency of each event is calculated by assuming 10 years before the final merger. The detectors (TianQin, LISA, LIGO A+, ET-D, and CE) are plotted
    separately. The GW signals are averaged over direction, inclination, and polarization. Note that for the triangle
    shaped detectors, the noise strains as shown should be multiplied by ${\rm
    sin}^{-1}60^{\circ}$. Furthermore, the average
    factors are absorbed in the signals instead of the noise strains.}
\label{Fig:WaveformSensitivityCurves}
\end{figure}

In Fig. \ref{Fig:WaveformSensitivityCurves}, we plot the characteristic strains of compact binary coalescence
from quasicircular binaries, with parameters derived from transient catalogs published by the LIGO-Virgo-KAGRA collabortion, 
together with the noise amplitudes listed detectors.
It can be observed that although GW190521 merges in the higher bands, 10 years before the merger, the early inspiral GWs
have a
frequency of as low as $\sim 10^{-2}$Hz and locates above the sensitive band of TianQin.
We further notice that due to the higher mass, GW190521, which is indicated with the black thick line, can extend to 
a lower frequency, making it easier to detect with TianQin.

The strength of a GW signal in a detector can be characterized by the signal-to-noise ratio (SNR). For one Michelson
interferometer, the SNR accumulated in observation time is as follows \cite{Cutler:1994ys}
\begin{align}\label{Eq:SNRofOneInterferometer}
  \rho = \sqrt{4\int_{f_{i}}^{f_{f}}\frac{\tilde{h}^{*}(f)\tilde{h}(f)}{S_{n}(f)}df},
\end{align}
where $\tilde{h}$ is the GW waveform in frequency domain , $``*"$ denotes the complex conjugate, $S_{n}(f)$ is the PSD
of the interferometer, $f_{i}$ and $f_{f}$ are the initial and final GW frequenciesi, respectively, which can be obtained by \cite{Cutler:1994ys}
\begin{align}\label{Eq:FofT}
  f(t) = (5/256)^{3/8}\frac{1}{\pi}\mathcal{M}^{-5/8}(t_{c}-t)^{-3/8},
\end{align}
where $\mathcal{M}=(m_{1}m_{2})^{3/5}/(m_{1}+m_{2})^{1/5}$, $t_{c}$ and $t$ are chirp mass, coalescence time, and
observation time, respectively.  
Note that when the observation time is equal to the coalescence time, $t=t_{c}$, we truncate the final frequency $f_{f}$
at the inner stable circular orbit $f_{\rm ISCO}=(6^{3/2}\pi M)^{-1}$ \cite{Cutler:1994ys}, with $M=m_{1}+m_{2}$ being the total mass. 
For the case with a signal observed by multiple interferometers simultaneously, the total SNR is
\begin{align}\label{Eq:SNRSummed}
  \rho = \sqrt{\sum_{j=1}^{n}\rho_{j}^{2}},
\end{align}
where $\rho_{j}$ is the SNR of a $j$th interferometer and $n$ is the total number of interferometers.
In order to comprehensively explore the observation potential of future space-borne GW detectors for GW190521-like events, we not only consider TQ and TQ I+II but also a joint observation of TianQin and LISA, i.e., TQ+LISA and TQ I+II+LISA.
The response functions and orbital motions for TianQin and LISA are described in detail in \cite{Liu:2020eko, Berti:2004bd} respectively.

\begin{figure}[h]
\centering
\includegraphics[width=0.5\textwidth]{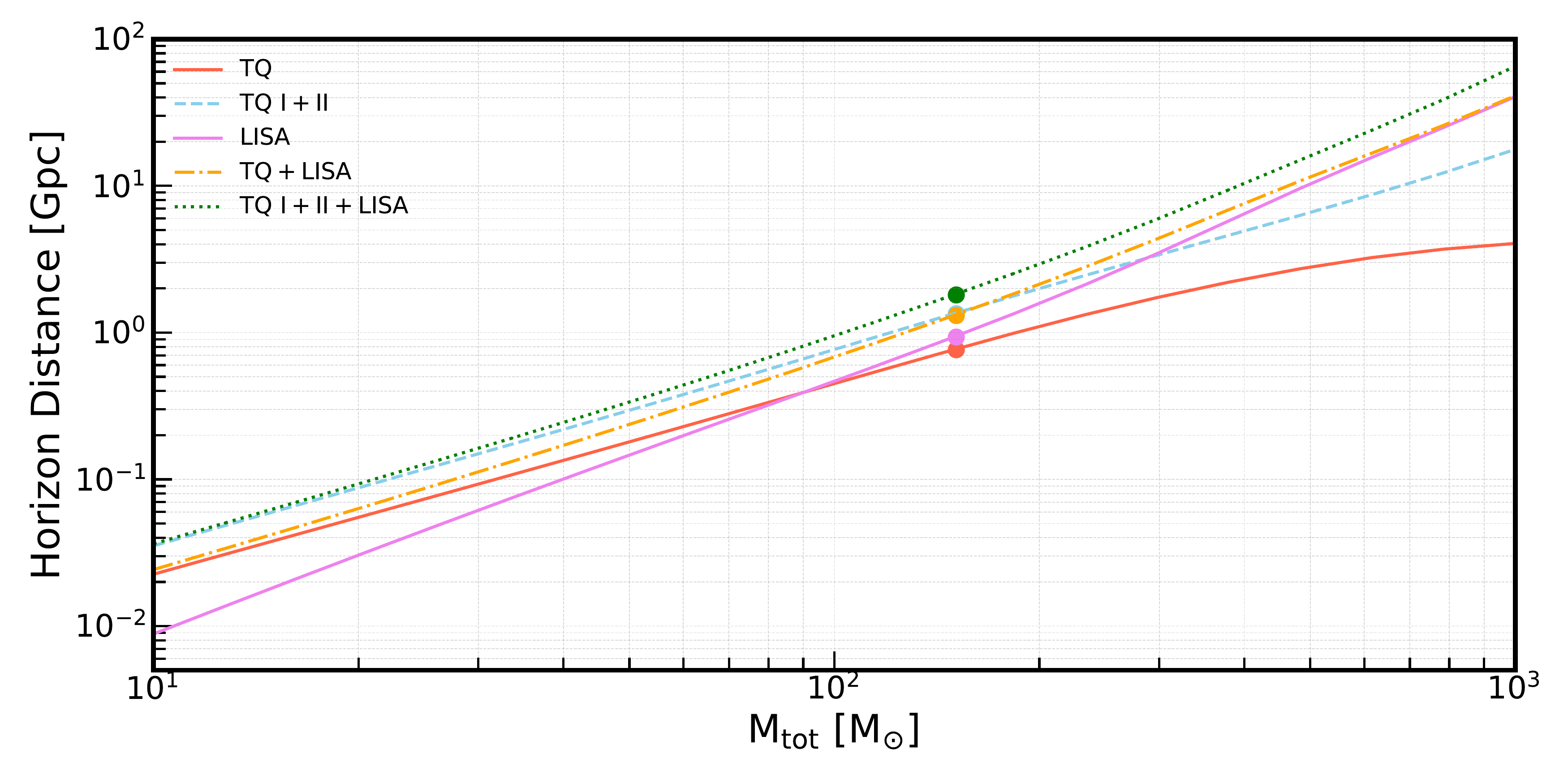}
    \caption{The horizon distance with the SNR threshold of 8 is plotted against
    the total mass of binary black holes.
    The different color lines represent TQ, TQ I+II, LISA, TQ+LISA, and TQ I+II+LISA, respectively. 
    All lines are plotted assuming equal mass binaries, and adopting an antenna response that averages over sky localization, inclination and polarization.
  The dots mark GW190521. We assume the operation time of TianQin and LISA is five and four years, respectively.}
\label{Fig:HorizonDistance}
\end{figure}

For a binary black hole, the horizion distance of a GW detector, i.e., the maximum detectable distance, could be
obtained by solving the Eq. (\ref{Eq:SNRofOneInterferometer}) under a given SNR threshold.
In Fig. \ref{Fig:HorizonDistance}, we show the horizon distance of TianQin for the equal mass binary black hole inspirals
averaged over sky localization, inclination, and polarization with a
total mass between $10-1,000\ M_{\odot}$, assuming an SNR threshold $\rho_{\rm thr}=8$ for detection.
We can see that the horizon distance of TQ grows with the increase of the total masses, because the GWs emitted by
heavier sources are stronger, their SNRs are larger. For GW190521-like binaries with a total mass of $\sim 150M_\odot$,
TianQin could reach a distance of as far as $0.7$Gpc.

It is also noted that the horizon distance of TQ increases slowly at the high mass tail, the reason is that for more
massive systems, the observation gaps from the ``3months on + 3months off" pattern \cite{Liu:2020eko} hinder the SNR accumulation.
When TQ I+II is considered, the horizon distance would become about twice that of TQ and show a linear growth, because TQ I+II has no observation gaps.
For LISA, the horizon distance is roughly the same as that of TQ I+II, with LISA being more sensitive for binaries with
the total mass higher than $\sim 100M_\odot$, but less sensitive when the total mass is smaller than $\sim 100M_\odot$. 
For TQ+LISA and TQ I+II+LISA, the horizon distance would increase, of which the latter shows more gain. For example, it
would be $\sim2\ {\rm
Gpc}$ for GW190521, because the latter contains the most detectors.
Furthermore, if we consider a further layer of the hierarchical merger model on top of two GW190521-like binaries, i.e.,
an equal mass binary with the total mass $\sim300M_{\odot}$, it could be
shown that TianQin (TianQin+LISA) would make detections for sources up to $\sim3\ (6){\rm Gpc}$. 

\begin{figure}[h]
\centering
\includegraphics[width=0.5\textwidth]{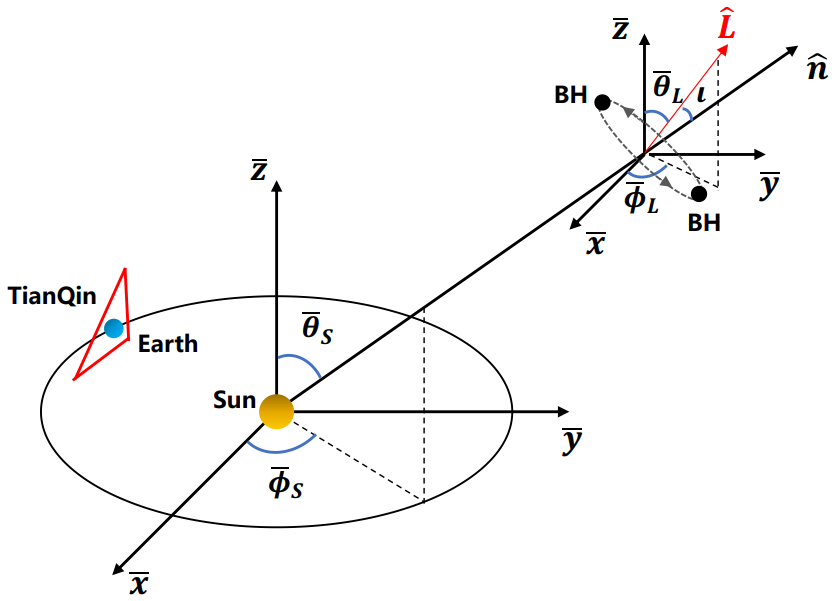}
    \caption{The schematic diagram of the TianQin observatory and a source, which are represented by a red triangle and two black dots. In the ecliptic coordinate system ($\bar{x}, \bar{y}, \bar{z}$), the source position and orbital angular momentum unit vectors $\hat{n}$ and $\hat{L}$ of a source are
    characterized by
    ($\bar{\theta}_{S}, \bar{\phi}_{S}$) and ($\bar{\theta}_{L}, \bar{\phi}_{L}$), respectively. $\iota$ is the inclination angle between them.}
\label{Fig:SourceSchematicDiagram}
\end{figure}

We further examine the detection number of GW190521-like sources in the range within redshift $z=2$.
According to the GW190521-like binary merger rate reported by the LIGO-Virgo collaboration \cite{LIGOScientific:2020iuh},
we reconstructed the probability density distribution (PDF) by fitting through a log-normal distribution, and
200 random realizations are generated according to this distribution. 
In each realization, the component masses are set to the median values estimated from GW190521 data. 
The angular parameters as described in Fig. \ref{Fig:SourceSchematicDiagram} are distributed uniformly on a sphere.
Which means ${\rm cos}\bar{\theta}_{S}$ and ${\rm cos}\bar{\theta}_{L}$ follow a uniform
distribution ${\rm U}[-1, 1]$, while $\bar{\phi}_{S}$ and $\bar{\phi}_{L}$ obey ${\rm U}[0, 2\pi]$. The overline denotes
that we adopt the parameters within the ecliptic coordinate system.

\begin{figure}[h]
\centering
\includegraphics[width=\textwidth]{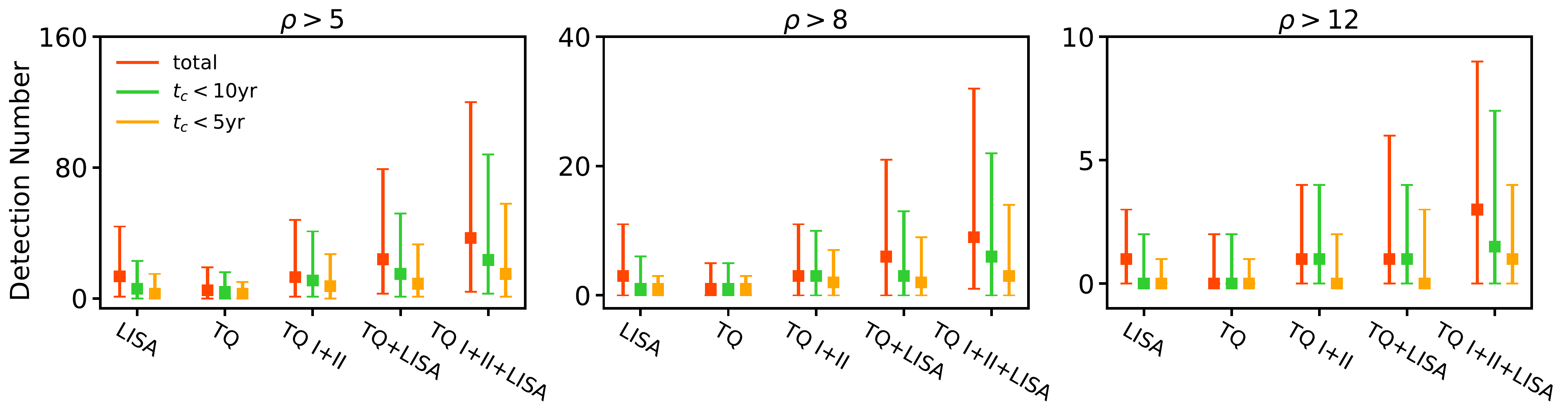}
    \caption{The detection number for GW190521-like binaries.
  The left, middle, and right panels correspond to an SNR threshold of 5, 8, and 12, respectively. Different detectors/networks are shown separately.
  For each configuration, we plot the 90\% confidence interval of the expected detection number within the operation
    time, different lines indicate events that will merge within 5, 10, or more years. The operating time of TianQin and
    LISA are five and four years respectively.}
\label{Fig:DetectionNumber}
\end{figure}

We present the expected detection number over a 5 yr operation time in Fig. \ref{Fig:DetectionNumber}.
With the SNR threshold of TianQin of 12 \cite{Liu:2020eko}, TQ or LISA could only make detections in the optimistic scenario.
The formation of a network through a number of detectors would improve the detection
ability. The expected detection
number increases with the order of TQ I+II, TQ+LISA, and TQ I+II+LISA, noticing that a network of TianQin and LISA could
detect up to $\sim10$ sources.
Notice that the coincident observation of multiple detectors could debunk a lot of false alarms, potentially bringing down
the SNR threshold \cite{Maggiore:2007ulw}. Therefore, we also consider the SNR threshold of 8.
In the most optimistic scenario, as many as $\sim30$ events could be detected.
It has been proposed that future ground-based GW detectors can detect SBBH mergers, and trigger a targeted search in space-borne GW detector archive data.
Such archival search will be performed in a shrunk parameter space, gaining the potential of further lowering the
detection threshold for SNR 5 \cite{Wong:2018uwb, Ewing:2020brd}. In such a case, up to $\sim120$ sources could be observed with TianQin.

It should be noted that about half of the above sources would merge into ground-based GW detector bands within 10 years,
which allows the multiband GW observation.
We discuss the parameter estimation precision on these sources later.

\subsection{Parameters estimation precision}\label{SubSec:Parameters_estimation}

Assuming the noise to be Gaussian and stationary, for an unbiased estimator of the physical parameters, one can estimate
the precision by looking at the covariance matrix, which can be derived with the Fisher information matrix (FIM) method.
For one interferometer, the FIM matrix is as follows \cite{Cutler:1994ys}
\begin{equation}
  \Gamma_{ij}=\left(\frac{\partial h}{\partial\lambda^{i}}\middle|\frac{\partial h}{\partial\lambda^{j}}\right),
\end{equation}
where $h(\lambda^{i})$ is the GW waveform determined by the parameter set $\lambda^{i}$.
For a network of $n$ interferometers, the total FIM is the summation of individual components 
\begin{equation}\label{Eq:FIMSummed}
  \Gamma_{ij}=\sum_{k=1}^n\Gamma_{ij}^{k},
\end{equation}
where $\Gamma_{ij}^{k}$ is the FIM of $k$th interferometers. The root mean square of the standard deviation of the $i$th parameter $\lambda^i$, i.e., the estimation
precision, is the square root of the variance, or the $ii$ component of the covariance
matrix $\Sigma$, which relates with the FIM $\Gamma$ through $\Sigma=\Gamma^{-1}$.

We apply the FIM estimate to simulated events with a SNR larger than 8 and will merge within five years, so that an early warning is possible and meaningful.
The parameters of sources we consider are $\bm{\lambda}=\{t_{c}, \bar{\theta}_{S}, \bar{\phi}_{S}, \mathcal{M}, \eta,
D_{L}, \iota, e_{0}\}$.
Since the eccentricities of sources at $\sim0.01$Hz where TianQin is most sensitive are generally lower than 0.1, so we
choose $e_{0}=0.01$ at 0.01Hz as a fiducial value \cite{Nishizawa:2016jji}.
The precision on the sky localization $\Delta\bar{\Omega}_{S}$ is \cite{Berti:2004bd}
\begin{equation}
  \Delta\bar{\Omega}_{S}=2\pi|\sin\bar{\theta}_{S}|(\Sigma_{\bar{\theta}_{S}\bar{\theta}_{S}}\Sigma_{\bar{\phi}_{S}\bar{\phi}_{S}}-\Sigma_{\bar{\theta}_{S}\bar{\phi}_{S}}^{2})^{1/2},
\end{equation}
and the error volume of a source is \cite{Liu:2020eko}
\begin{equation}\label{Eq:DeltaV}
  \Delta V\sim D_{L}^{2}\Delta\bar{\Omega}_{S}\Delta D_{L}.
\end{equation}

\begin{figure}[h]
\centering
\includegraphics[width=\textwidth]{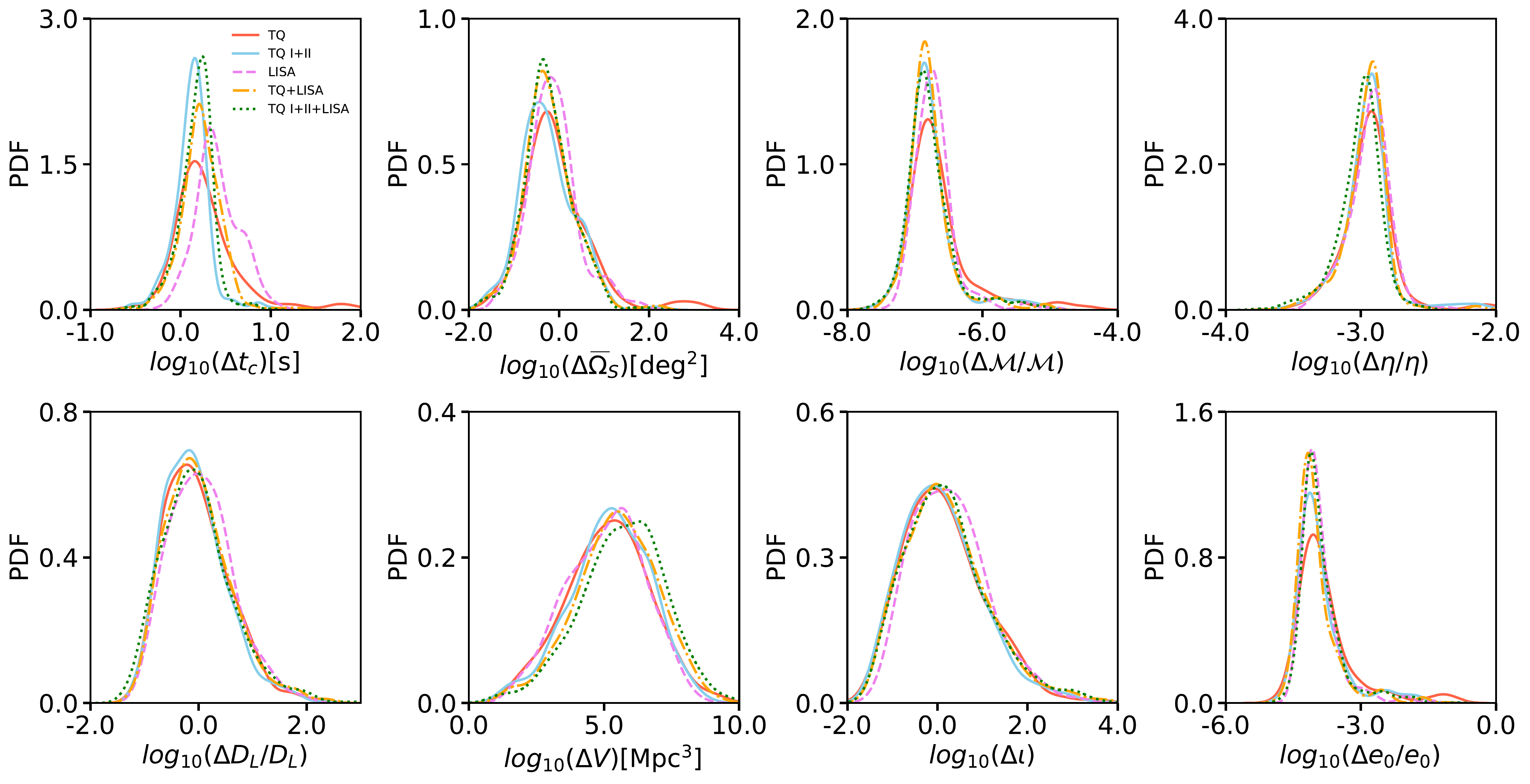}
    \caption{Parameter estimation precision for GW190521-like events with SNR greater than 8 and
    merging in five years. The parameters include coalescence time $t_{c}$, sky localization $\bar{\Omega}_{S}$, chirp mass
    $\mathcal{M}$, symmetric mass ratio $\eta$, luminosity distance $D_{L}$, error volume $\Delta V$, inclination angle
    $\iota$, and orbital eccentricity $e_{0}$. The different color and style lines denote TQ, TQ I+II, LISA, TQ+LISA,
    and TQ I+II+LISA, respectively.}
\label{Fig:ParaEstiPrecDist}
\end{figure}

The precision distributions on these parameters are shown in Fig. \ref{Fig:ParaEstiPrecDist}. 
For TQ, the precision of coalescing time and sky localization is of the order $\sim1\ {\rm s}$ and $\sim 1\ {\rm deg^{2}}$,
respectively.
Such high precision in sky localization is achieved through the modulation of the TianQin's orbital motion \cite{Liu:2020eko}.
We remark that this precision is vital to the multiband GW observation and multimessenger observation.

In addition, the mass parameters in the waveform phases could also be recovered very accurately, to a relative precision
of $10^{-7}$. This is because the phase evolution is very sensitive to intrinsic parameters like mass, and the precision is inversely proportional to the number of cycles observed, or $\propto 1/N_{\rm cyc}$ \cite{Cutler:1994ys}.
For the other mass parameter, such as the symmetric mass ratio, it also affects the evolution of the waveform phase. However,
it is only on higher order terms, so its relative precision is worse, but still could reach $\sim10^{-3}$.

In comparison, the current ground-based GW observation can only constrain the mass to the accuracy of
$\mathcal{O}(0.1)$, which is also prior dependant, with some studies conclude very different mass estimations
\cite{LIGOScientific:2020iuh, Fishbach:2020qag}.
The space-borne GW detection could easily erase such uncertainty and pinpoint the mass, which would also be important for reconstructing the underlying mass function if multiple events are detected.

For a BBH system, the luminosity distance $D_L$ could be determined to a relative precision of $\sim1/\rho$\cite{Cutler:1994ys}.
With the SNR threshold of $8$, we expect a relative uncertainty of 10\% for the luminosity distance.
Combining the luminosity distance and sky localization, one could deduce the error volume with Eq. (\ref{Eq:DeltaV}).
Although the scarcity of the events makes most sources far away and associated with small SNR, both indicators for
larger error volumes, one could still expect exceptions of relatively nearby events, with error volumes as small as
$100\ {\rm Mpc^{3}}$.
On the other hand, the average number density of Milky Way equivalent galaxies is $\sim0.01\ {\rm Mpc^{-3}}$
\cite{Kyutoku:2016ppx}, which means that one could expect to pinpoint the host galaxy through the GW observation. We discuss the implication of this host galaxy identification on GW cosmology later.

We notice that although the initial orbital eccentricity is set to be a very small number of $\sim0.01$, TianQin could
still observe it with very high precision, with a relative uncertainty of $\sim10^{-4}$. This precise measurement
ability on orbital eccentricity would almost definitely determine the formation channel of such GW190521-like binaries.

Finally, it should be noted that for the networks composed of multiple detectors (TQ I+II, TQ+LISA, and TQ I+II+LISA), the results are
similar to those from TQ or LISA. This is because for whatever configurations adopted, the SNR threshold is kept fixed,
and the distribution for detected events depends only on the threshold. More detectors are helpful only through the
sense that more events can be observed and the loudest event is expected to be associated with higher SNR.

\section{The capability to constrain the Hubble constant}\label{Sec:ConstrainHubbleConstant}

In this section, we explore the potential of TianQin on constraining cosmology through observation of GW190521-like events.
Throughout the work, we assume the underlying cosmology to follow the standard $\Lambda$CDM model, where the Hubble parameter can be expressed as 
\begin{equation}\label{Eq:H_z}
H(z) = H_0 \sqrt{\Omega_M(1+z)^3 + \Omega_{\Lambda}},
\end{equation}
with $H_0 = 67.8 $ km/s/Mpc and $\Omega_M = 0.307$ adopted \cite{Planck:2015fie},
where $z$ is the redshift, the Hubble constant $H_0 \equiv H(z=0)$ describes the current expansion rate of the Universe, and $\Omega_M$ and
$\Omega_{\Lambda} = 1 - \Omega_M$ are the fractional densities for total matter and dark energy with respect to the
critical density, respectively. The luminosity distance of a BBH in the Universe can be calculated with the
Hubble constant and its redshift. As mentioned above, the luminosity distance of a BBH could be determined with GW
observation;
if the redshift of the galaxy in which the source resides is known, then the Hubble constant could be constrained by fitting the
relationship between the luminosity distance $D_L$ and redshift $z$.

We adopt a Bayesian analysis method to infer the Hubble constant from GW190521-like GW detection data $d_{\rm GW}^i$ and assisting EM observation data $d_{\rm EM}^i$ \cite{Chen:2017rfc, Fishbach:2018gjp, Zhu:2021aat}. 
In such case, the likelihood can be written as
\begin{equation} \label{likeli_i}
p(d_{\rm GW}^i, d_{\rm EM}^i| \mathcal{H}, I) = \frac{\int p(d_{\rm GW}^i, d_{\rm EM}^i, D_L, z, \bar{\phi}_{S}, \bar{\theta}_{S},
    G, L | \mathcal{H}, I) \D D_L \, \D z \, \D \bar{\phi}_{S} \, \D\bar{\theta}_{S} \, \D G \, \D L}{\beta(\mathcal{H} | I)},
\end{equation}
where $i$ denotes the $i$th event detected, $\mathcal{H} \equiv \{H_0, \Omega_M \}$ represents the cosmological parameters set, $I$ represents all relevant
background information, $\bar{\phi}_{S}$, $\bar{\theta}_{S}$, and $L$ are the polar
angle, azimuthal angle, and luminosity,
respectively, $G$
denotes the galaxy hosting a source, and $\beta(\mathcal{H}|I)$ is the normalization coefficient. 

We can factorize the integrand of the numerator in Eq. (\ref{likeli_i}) as 
\begin{align} \label{likeli_1}
p(d_{\rm GW}^i, d_{\rm EM}^i, D_L, z, \bar{\phi}_{S}, \bar{\theta}_{S}, G, L | \mathcal{H}, I) = \ & p(d_{\rm GW}^i | D_L,
    \bar{\phi}_{S}, \bar{\theta}_{S}, I) p(d_{\rm EM}^i | z, \bar{\phi}_{S}, \bar{\theta}_{S}, L, I)  \nonumber  \\
& \times p_0(D_L | z, \mathcal{H}, I) p_0( G | L, \mathcal{H}, I) p_0(z, \bar{\phi}_{S}, \bar{\theta}_{S}, L |
    \mathcal{H}, I),
\end{align}
where $p_{0}$ represents the prior. In this work, we work under the dark standard siren scenario, where we assume no direct observation data of EM
counterpart, therefore we can set $p(d_{\rm EM}^i | z, \bar{\phi}_{S}, \bar{\theta}_{S}, L, I) = {\rm constant}$ \cite{Chen:2017rfc, Fishbach:2018gjp}. 
We define the prior $p_0(D_L | z, \mathcal{H}, I) \equiv \bar{\theta}_{S}( D_L - \hat D_L(z, \mathcal{H}) )$ which is
sensitive to the cosmological model, where $\hat{D}_{L}(z, \mathcal{H})$ is the functional relationship between
redshift and luminosity distance \cite{Hogg:1999ad}. 
The brighter galaxies generally contain more compact objects, and we assume that the probability of a galaxy hosting a
GW190521-like binary is proportional to its $K$-band luminosity, then $p_0( G | L, \mathcal{H}, I) \propto L_K(G)$ \cite{Fishbach:2018gjp, Gray:2019ksv, Abbott:2019yzh, Vasylyev:2020hgb}. 
Since the horizon distances of GW190521-like events by both TianQin and LISA are only about $1 \ \rm Gpc$ \cite{2012RAA....12.1197C, Aghamousa:2016zmz, Gong:2019yxt, Gardner:2006ky},
therefore, we assume the error in the EM survey measurements in our calculations is small and can be safely ignored, $p_0(z, \bar{\phi}_{S}, \bar{\theta}_{S}, L | \mathcal{H},
I) \propto \sum\nolimits_{j} \delta (z - z_j) \delta (\bar{\phi}_{S} - \bar{\phi}_{Sj})
\delta (\bar{\theta}_{S} - \bar{\theta}_{Sj}) \delta (L - L_j)$, where $j$ denotes $j$th event
detected (in this work, we adopt a mock galaxy catalog from the MultiDark Planck $N$-body cosmological simulation, obtained from the
Theoretical Astrophysical Observatory \footnote{\href{https://tao.asvo.org.au/tao/}{https://tao.asvo.org.au/tao/}}
\cite{Klypin:2014kpa, Croton:2016etl, Conroy:2009gw}). 
Note that the redshift errors caused by the peculiar velocities of galaxies are taken into account in our cosmological
analysis, and they are equivalently translated into an additional error of $D_L$ to the GW source in the calculation
processes \cite{He:2019dhl}. 

The normalization term $\beta(\mathcal{H} | I)$ can be used to account for selection effects and ensure that the likelihood integrates to unity \cite{Chen:2017rfc, Mandel:2018mve}. 
In this work, we follow the statistical method presented in the literature \cite{Zhu:2021aat} to evaluate the selection
biases of the survey galaxies catalog and calculate the normalization term. 

We adopt two methods to count the statistical redshift distribution of candidate host galaxies of GW source, as follows 
\begin{enumerate}[i.]
  \item fiducial method: each galaxy in the spatial localization error box of the GW source has equal weight regardless
      of its position and luminosity information.
  \item weighted method: the weight of a galaxy is related to both its position and $K$-band luminosity.
\end{enumerate}  

\begin{figure}[h]
\centering
\includegraphics[width=0.8\textwidth, height=0.5\textwidth]{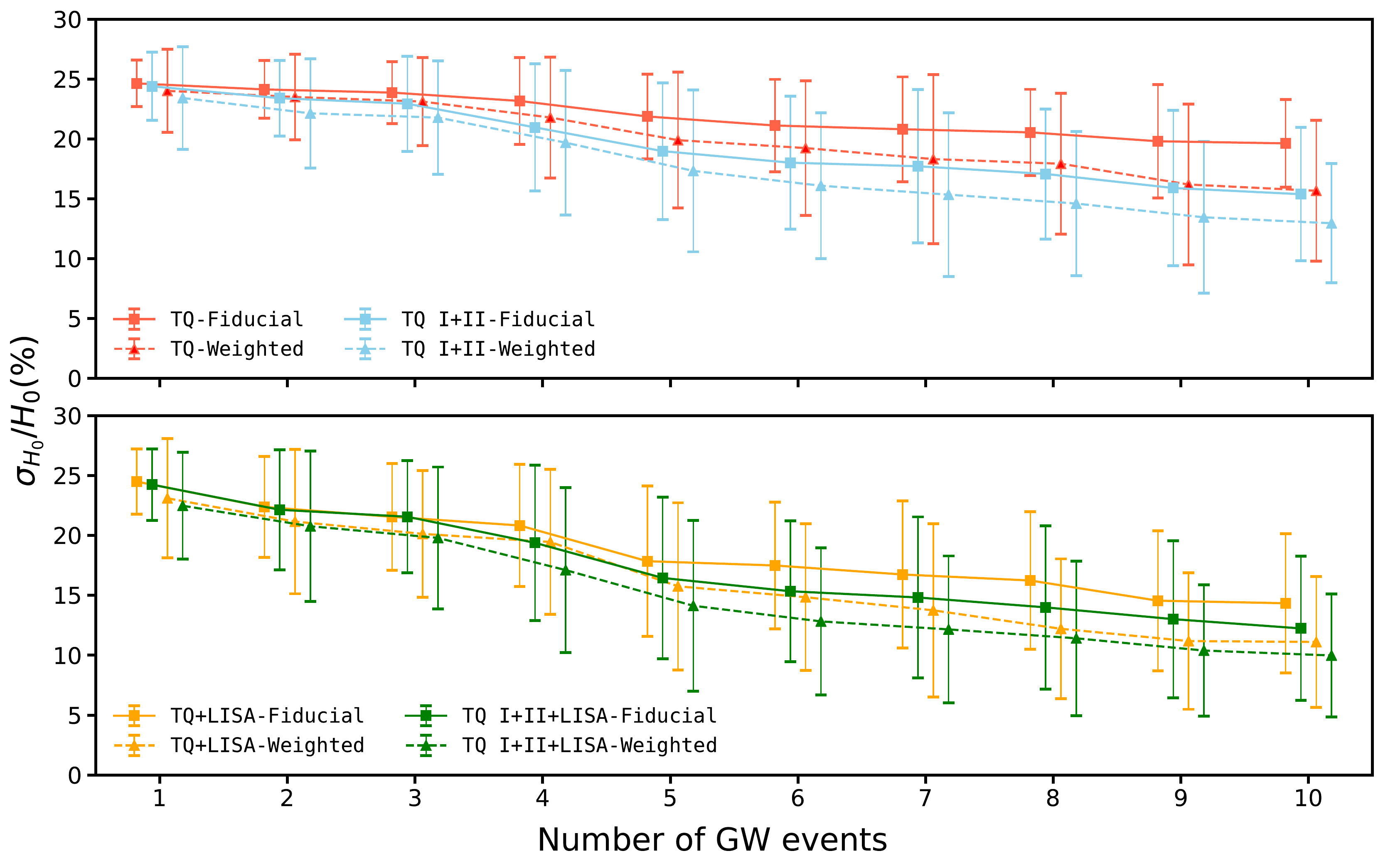}
  \caption{Dependence of constraint precision of the Hubble constant (with 68.28\% credible intervals) on numbers of GW
    events. 
  The upper panel shows the constraint results of TianQin (red) and TQ I+II (cyan); 
  the lower panel shows the constraint results of TQ+LISA (orange) and TQ I+II+LISA (green). 
  Solid lines and dashed lines represent fiducial and weighted methods, respectively.}
\label{Fig:H0_error}
\end{figure}

We adopt the affine-invariant Markov chain Monte Carlo ensemble sampler EMCEE \cite{ForemanMackey:2012ig, ForemanMackey:2019ig} to perform the cosmological parameter estimation. 
Since the detection numbers of GW190521-like events are very uncertain (see Fig. \ref{Fig:DetectionNumber}), we
demonstrate the capability of GW190521-like events to constrain the Hubble constant in the form of the evolution of the
constraint precision with the number of GW events, as shown in Fig. \ref{Fig:H0_error}. 
In order to eliminate the random fluctuations caused by the specific choice of any event, we repeat this process 48 times
independently for each number of GW events, and using the mean value of the constraint errors and $68.27\%$ credible interval to plot the error bars. 
Due to the computation limit, we only adopt the GW sources with $\Delta D_L / D_L < 1$ and $\Delta \bar \Omega_S < 100\ {\rm
deg}^2$ use to the analyses of constraining the Hubble constant, because the discarded events can provide little
improvements on cosmology constraints. 

If we assume five GW190521-like events detected, then TQ could constrain the Hubble constant with precisions of about $22\%$
and about $20\%$, using the fiducial method and weighted method, respectively. 
Compared with TQ, TQ I+II has a better capability to constrain the Hubble constant. 
Under the condition of the same number of GW events, the precision of the Hubble constant constrain from TQ I+II using
fiducial method slightly outperformed TianQin using the weighted method. 
With five events, TQ I+II could constrain the Hubble constant with precisions of about $19\%$ and $17\%$ using the fiducial method and the weighted method, respectively. 

For the network of multiple space-borne GW detectors, such as TQ+LISA or TQ I+II+LISA, 
it could significantly improve the capability of the same GW events of constraining the Hubble constant. 
If we consider the most optimistic scenario where $10$ GW190521-like events are detected, for the TQ+LISA configuration,
the Hubble constant are expected to be constrained to precisions of about $14\%$ and $11\%$ using the fiducial method
and the weighted method, respectively. For the TQ I+II+LISA configuration, the constraint precision of the Hubble constant is expected to achieve the level of
about $10\%$ using the weighted method. 

\begin{figure}[h]
\centering
\includegraphics[height=7.5cm, width=7.5cm]{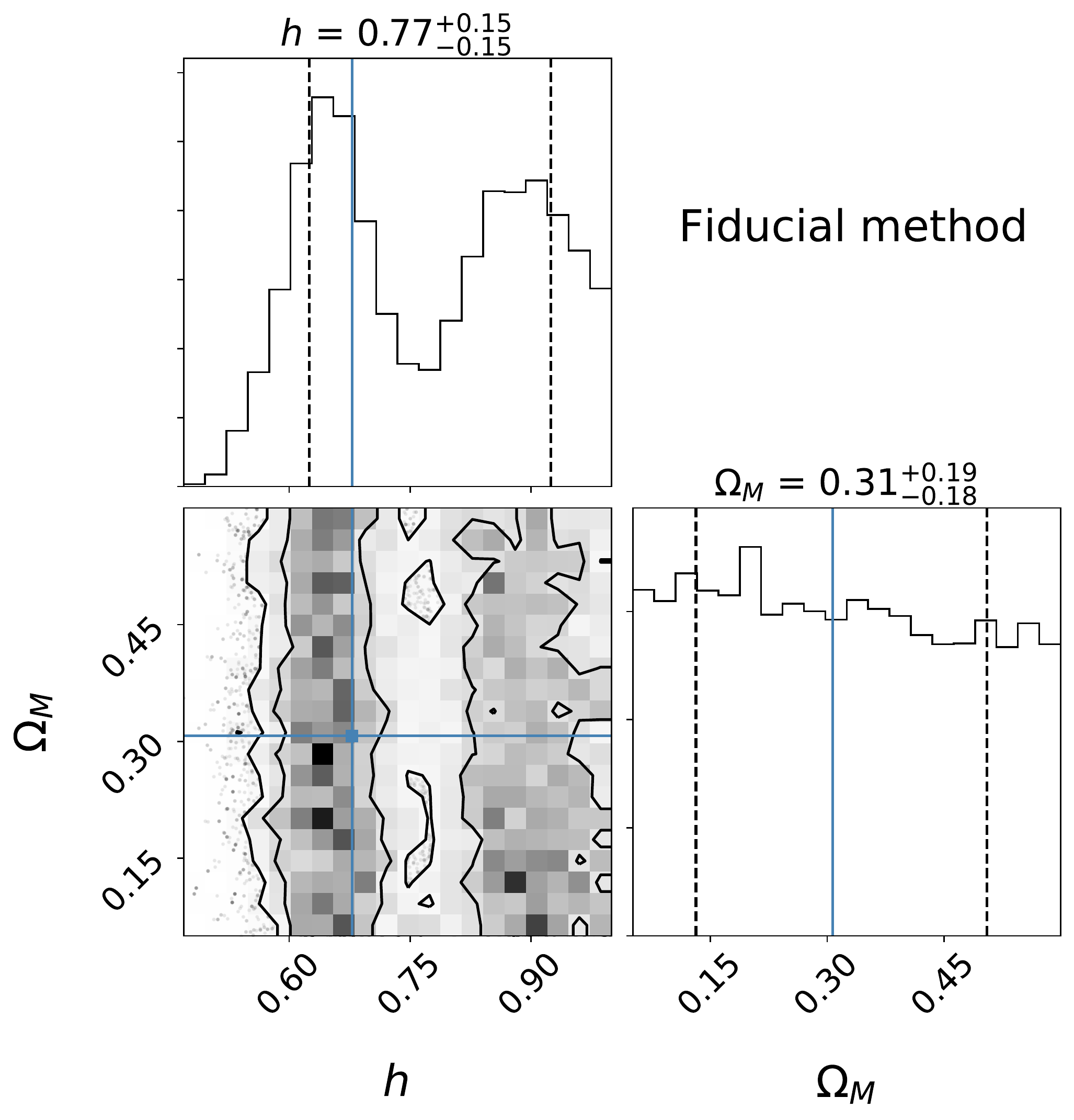}
\includegraphics[height=7.5cm, width=7.5cm]{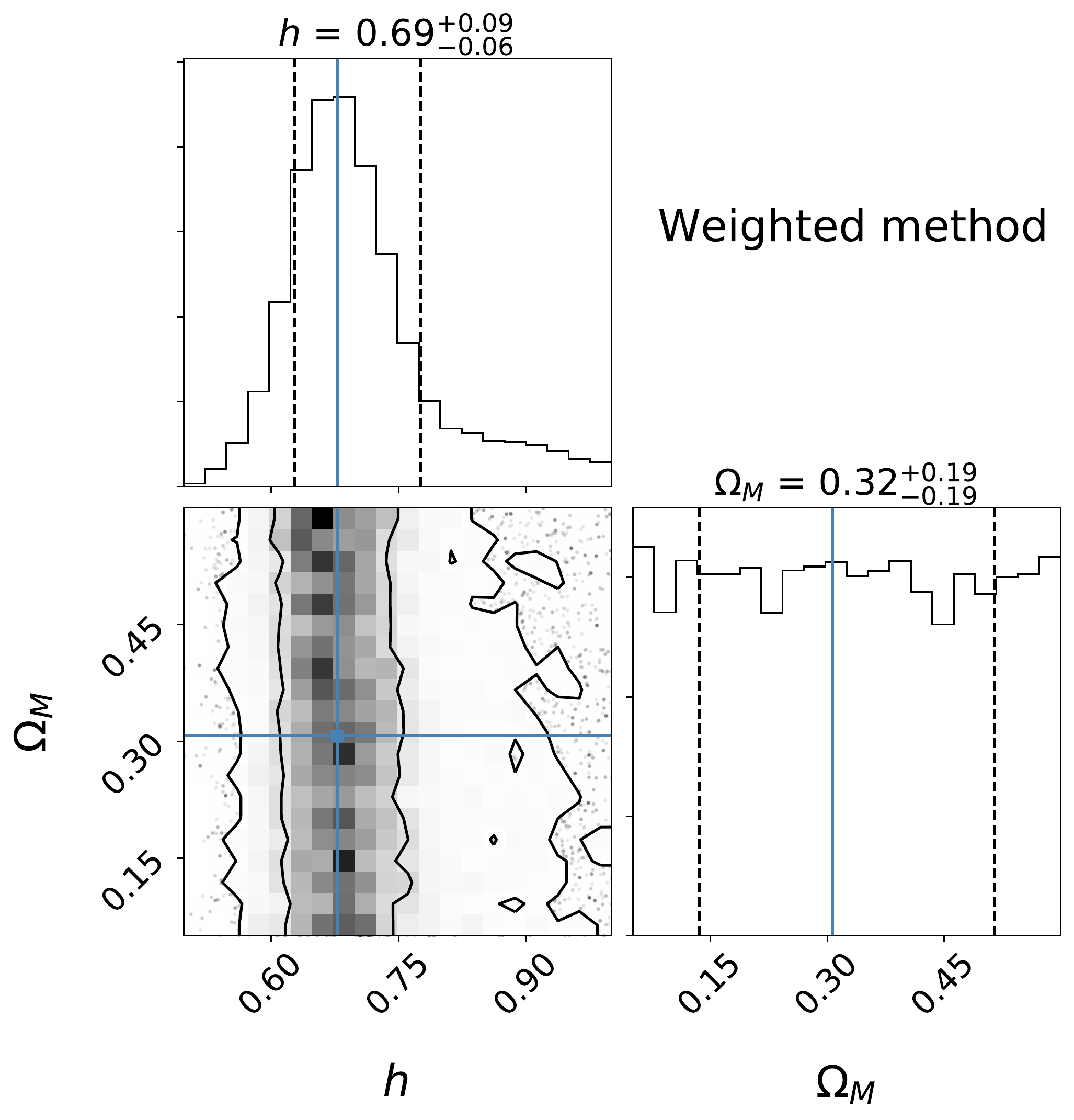}
    \caption{Examples of the posterior probability of the parameters $h$ ($h \equiv \frac{H_0}{100 {\rm km/s/Mpc}}$) and
    $\Omega_M$ using fiducial method (left plot) and weighted method (right plot), constraints from the same six GW events
    detected with TianQin. 
  In each plot, the lower left panel shows the joint posterior probabilities of $h$ and $\Omega_M$, the contours represent confidence levels of $1 \sigma (68.27\%)$ and $2\sigma (95.45\%)$, respectively; the upper and right panels show the histogram of the posterior probability of same parameters after marginalized the other one parameter, while the dashed lines indicate $1 \sigma$ credible interval. 
  In each panel, the cyan lines mark the injected parameters.}
\label{Fig:H0_example_1TQ}
\end{figure}

In various detector configurations, compared with the fiducial method, the weighted method could consistently improve the estimation precision of the Hubble constant. To demonstrate the effect of the weighted method on the constraint of the Hubble constant, we show an example of
cosmological parameter estimation using the fiducial method and the weighted method, respectively, in Fig.
\ref{Fig:H0_example_1TQ}. 
Notice that for the fiducial method, the contamination of galaxies other than the host can cause multiple peaks in the posterior, leading to a worse precision.
On the other hand, the weighted method can reliably associate higher weights to the correct host, therefore shrinking the uncertainties.

%%%--------------------------------------------------------------------
\section{Discussion}\label{Sec:Discussion}

\begin{figure}[h]
    \centering
\includegraphics[width=0.5\textwidth]{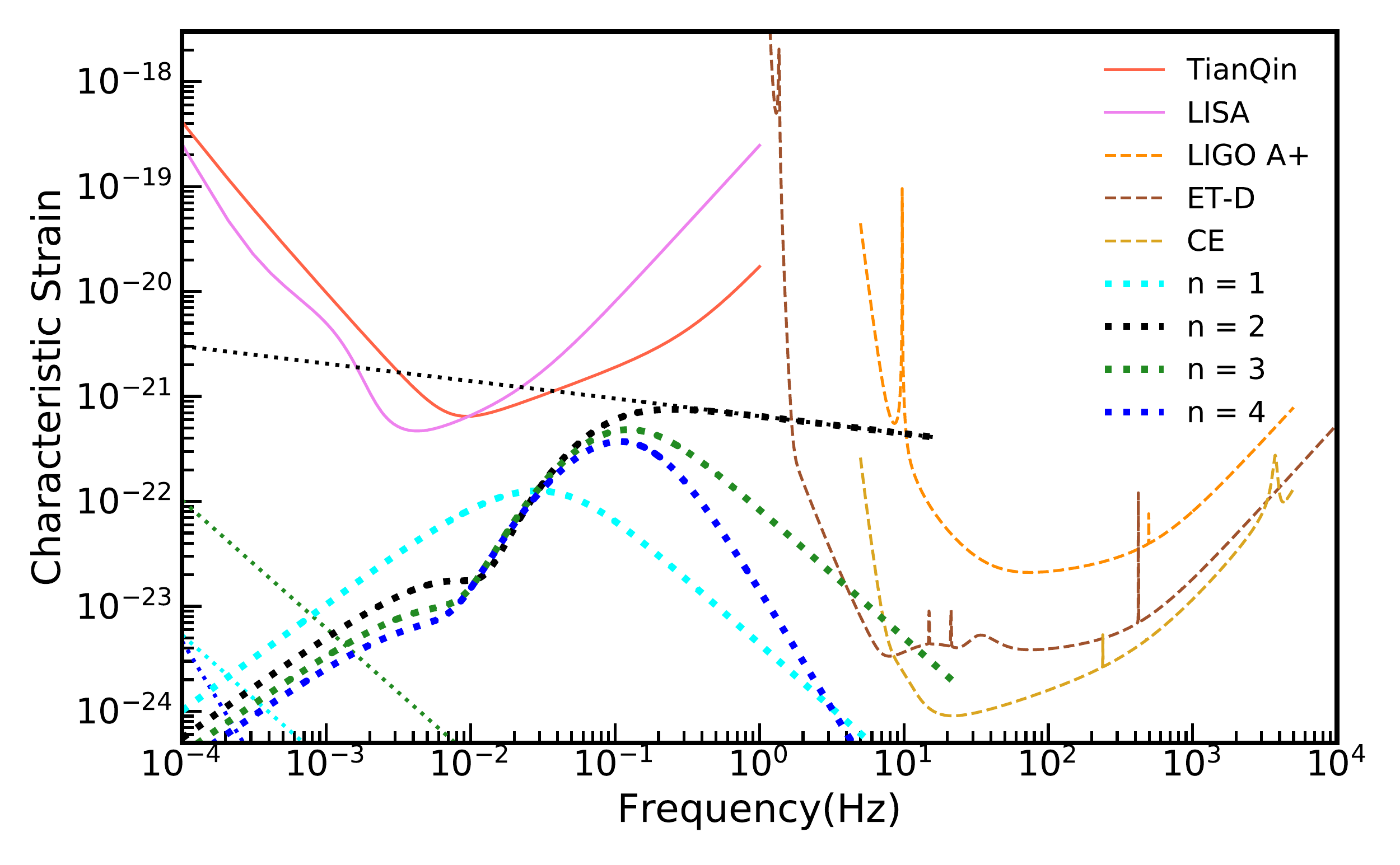}
    \caption{The characteristic strains of first four harmonics emitted by a source with initial eccentricity of $e_{0}=0.04$
    ($0.996$) with thin (thick) line at initial semimajor axis of 0.08 A.U. \cite{Peters:1963ux, Peters:1964zz,
    Barack:2003fp}. We inherit the parameters like mass and distance from GW190521. The GW signals are averaged over
    direction, inclination, and polarization.}
\label{Fig:WaveformEccentric}
\end{figure}

In the previous section, we discuss GW190521-like binaries with small orbital eccentricities.
However, some theoretical models predict GW190521-like binaries with significant orbital eccentricities could be formed
by the dynamical processes \cite{Romero-Shaw:2020thy, Bustillo:2020ukp, Gayathri:2020coq}.
In Fig. \ref{Fig:WaveformEccentric} we demonstrate how strongly the binary evolution is affected by very large orbital
eccentricities.

For the source with small orbital eccentricity, the $n=2$ harmonic is always the dominant mode. 
The detectability by TianQin or LISA has been demonstrated through the previous section as well as a number of studies
\cite{Barack:2003fp, Kremer:2018tzm, Randall:2018lnh, DOrazio:2018jnv, Arca-Sedda:2018qgq, Zevin:2018kzq, Chen:2017gfm, Kremer:2018cir, Banerjee:2020cen}. 
However, for the system with very large eccentricity, the $n=2$ harmonic no longer dominates. With the radiation of GWs, the orbital is gradually
circularized, $n=2$ harmonic dominates the GW emission at high frequencies. 
We remark that for such sources, the GW strain they emit at the low frequency range could be too weak for either TianQin
or LISA to observe.
On the other hand, although the future generation ground-based GW detectors could detect them, with SNR of $\sim657$ and
$\sim371$ for CE and ET-D, respectively \cite{Barack:2003fp, Enoki:2006kj}, it would lose the ability to constrain the eccentricity due to the circularization.
The expected SNR for space-borne GW detectors is generally small, but one could expect to find the signal through the archival search triggered by ground-based detectors alerts \cite{Ewing:2020brd}. 
Therefore, a null observation in space-borne GW missions could indicate a very high eccentricity, which still
contributes significantly to revealing the formation mechanism of GW190521-like binaries.

\begin{figure}[h]
\centering
\includegraphics[width=0.5\textwidth]{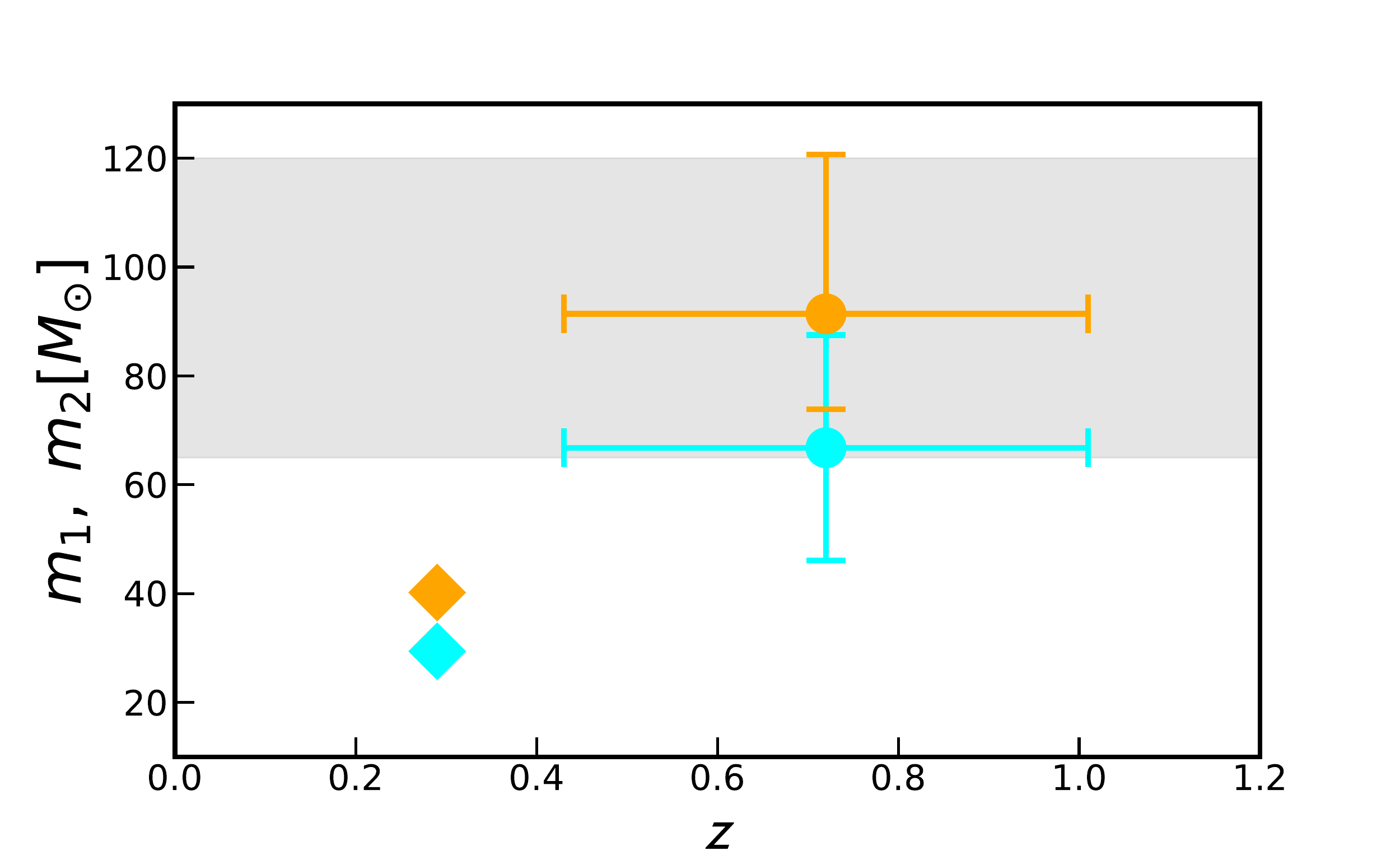}
    \caption{A schematic diagram of possible environment effects on GW190521, which could contribute additional redshift and lead to overestimation to mass parameters by a factor of as large as 3. The apparent (true) component masses and redshift are represented by
    circles (diamonds).}
\label{Fig:Overweight}
\end{figure}

If the LIGO/Virgo estimation of the physical parameters of GW190521 is correct, then future generation ground-based detectors
can certainly detect them.
However, environment effects could bring a shift to the estimated parameters \cite{Chen:2020iky}. 
For example, if GW190521 orbits around an SMBH, the relative motion between it and the SMBH would cause additional Doppler redshift
$z_{\rm dop}$ and gravitational redshift $z_{\rm gra}$. Such effects could bring a bias in the redshifted mass by a factor
$(1+z_{\rm dop})(1+z_{\rm gra})$ of as high as 3 \cite{Chen:2020iky}. Under this extreme scenario, the component masses
of GW190521 would become as small as $40.2M_{\odot}$ and $29.4M_{\odot}$, which could avoid the mass gap issue as shown in Fig. \ref{Fig:Overweight}. 
Again, due to the short duration of the signal at high frequencies, ground-based detectors could not effectively resolve
degeneracy raised by such environment effects. While the long duration nature of space-borne detections could help track the evolution of redshift and decipher the environment \cite{Toubiana:2020drf}.

\section{Conclusion}\label{Sec:Conclusion}
In this
work, we explore the detection capability of TianQin for GW190521-like binaries with small
eccentricities, i.e., horizon distance, detection number, and parameter precision. In addition, we examine the
improvement from multiple detectors (TQ I+II, TQ+LISA, and TQ I+II+LISA) compared with single detector. We also discuss the
application potential of such sources to GW cosmology.

For GW190521-like binaries, the horizon distance of TianQin is $\sim1\ {\rm Gpc}$, and the joint observation of TianQin
and LISA could reach to $\sim2\ {\rm Gpc}$. By adopting an SNR threshold of 12, TianQin or LISA could detect a few events, while
the joint observation of TianQin and LISA would improve the detection number to about 10.
Lowering the threshold to eight doubles the expected detection number.
If we consider the scenario of merger-triggered archival search, then it is possible to further lower the threshold to
five,
where TianQin (TianQin+LISA) could detect dozens (up to a hundred) events.
It is worth noting that about half of such sources would merge in several years, making them ideal multiband candidates.

We use FIM to estimate parameter precision, considering multiband sources with SNR greater than 8 and will merge within
five years.
We conclude that TQ could accurately estimate the parameters, with the coalescence time and sky localization
determined to the precision of $1\ {\rm s}$ and $1\ {\rm deg^{2}}$, respectively. This implies that TQ could predict when and where these
sources would merge. Such potential of early warning could help get facilities prepared to achieve multiband GW
observations and multimessenger observations, bringing hope to provide a better test of GR and more detailed study on surrounding environments.
We deduce that the error volumes of some loud
sources could be smaller than $100\ {\rm Mpc^{3}}$, indicating a direct pinpointing of the host galaxies, which makes them
ideal sources to perform GW cosmology study and put constraints on the Hubble constant. The orbital eccentricities could be measured
to a relative precision of $\sim10^{-4}$, which could help us to determine the formation mechanisms: from dynamical interaction in dense stellar clusters or isolated binary evolution. Finally, the $\sim 10^{-7}$ relative precision for chirp mass brings hope for the measurement of the mass function of IMBHs merged from GW190521-like binaries.
In the typical realization, TQ or TQ I+II could detect about five GW190521-like events, the constraint precision of the
Hubble constant are around $20\%$ for TQ and TQ I+II. Both the weighted method and the detector network could
significantly improve the capability to constrain the Hubble constant. 
In the optimistic scenario, if the network detects 10 GW190521-like events, using the weighted method, the constraint on
the Hubble constant could reach a precision of $10\%$.

We also discuss the possibility of detecting very eccentric GW190521-like binaries with TianQin or LISA. 
Although TianQin or LISA might miss the very eccentric binaries, the null detections could still contribute to the
deeper understanding of the sources by putting stringent constraints on orbital eccentricities. TianQin could also help
to break possible degeneracy raised by environment effects. 

Furthermore, it should be noted that the sources whose SNRs are below a given SNR threshold would form self
noise \cite{Karnesis:2021tsh}, and it is not considered in our calculation. The formed noise would still be below the noise
sensitivity \cite{Karnesis:2021tsh, Liang:2021bde}, so we believe that our conclusions should be robust. We will study the implication of the self noise on the detectability in future.

\begin{acknowledgments}
This work has been supported by the fellowship of China Postdoctroral Science Foundation (Grant No. 2021TQ0389), the Natural Science Foundation of China (Grants No. 12173104, No. 11805286, No. 91636111, and No. 11690022), Guangdong Major Project of Basic and Applied Basic Research (Grant No. 2019B030302001). 
The authors acknowledge the uses of the calculating utilities of NUMPY \cite{vanderWalt:2011bqk}, SCIPY
    \cite{Virtanen:2019joe}, and EMCEE \cite{ForemanMackey:2012ig, ForemanMackey:2019ig}, and the plotting utilities of
    MATPLOTLIB\cite{Hunter:2007ouj}. 
  The authors also thank Chang Liu, Zheng-Cheng Liang, Xiang-Yu Lyu, Shun-Jia Huang, and Jian-Wei Mei for helpful discussions.

\end{acknowledgments}

%%%%%%%%%%%%%%%%%%%%%%%%%%%%%%%%%%%%%%%%%%%%%%%%%%%%%%%%%%%%%%%
\bibliography{reference}
%%%%%%%%%%%%%%%%%%%%%%%%%%%%%%%%%%%%%%%%%%%%%%%%%%%%%%%%%%%%%%%
\end{document}